\documentclass{iau} 
\usepackage{graphicx}

\title[IMBH Feedback in Dwarf Galaxies] 
{Intermediate-Mass Black Hole Feedback in Dwarf Galaxies: 
a View from Cosmological Simulations} 

\author[P. Barai \& E. M. de Gouveia Dal Pino]   
{Paramita Barai$^1$ \and Elisabete M. de Gouveia Dal Pino$^1$} 

\affiliation{ 
$^1$Instituto de Astronomia, Geof\'isica e Ci\^encias Atmosf\'ericas - 
Universidade de S\~ao Paulo (IAG-USP), 
Rua do Mat\~ao 1226, S\~ao Paulo, 05508-090, Brasil \\ email: {\tt paramita.barai@iag.usp.br} 
} 

\pubyear{2018}
\volume{342}  
\jname{Perseus in Sicily: from black hole to cluster outskirts}
\editors{K. Asada, E. de Gouveia dal Pino, H. Nagai, R. Nemmen \& M. Giroletti, eds.} 

\begin{document}

\maketitle 

\begin{abstract} 
Black holes are usually observed to be of stellar-mass or supermassive. 
By natural extension, there should be a population of Intermediate-Mass Black Holes 
(IMBHs: with mass between $100$ to $10^6 M_{\odot}$) in the Universe; 
which has started to been observed. 
An exciting claim has been made recently by \cite{Silk17}: 
that early feedback by IMBHs in gas-rich dwarf galaxies at $z=5-8$, can potentially 
solve multiple dwarf galaxy problems within the $\Lambda$-cold-dark-matter cosmology. 
We are performing Cosmological Hydrodynamical Simulations 
of $(2 Mpc)^3$ volumes, starting from $z=100$, to test the case for IMBHs in Dwarf Galaxies. 
Black holes of mass $1000 M_{\odot}$ are seeded inside 
halos when they reach a mass of $10^7 M_{\odot}$. 
The black holes grow by accretion of gas from their surroundings 
and by merger with other black holes, and consequently eject feedback energy. 
We analyze the simulation output in post-processing to study the 
growth of the first IMBHs, and their impact on star-formation. 
Our conclusions, based on numerical simulation results, 
support the phenomenological ideas made by \cite[Silk (2017)]{Silk17}. 
IMBHs at the centers of dwarf galaxies can be a strong source of feedback to 
quench star-formation and generate outflows. 
At the same time, these IMBHs form the missing 
link between stellar-mass and supermassive BHs. 
\end{abstract} 

\firstsection 
\section{Introduction} 

Black holes usually come in two flavours: stellar-mass $(M_{\rm BH} \leq 10 M_{\odot})$, 
and supermassive $(M_{\rm BH} \geq 10^{6} M_{\odot})$.
Naturally, there should be a population of Intermediate-Mass Black Holes (IMBHs)
of masses between $100 - 10^{6} M_{\odot}$.
Analogous to supermassive BHs producing AGN feedback, the IMBHs should also have feedback. 
In this work we focus on negative BH feedback effects where star-formation is quenched. 

AGN feedback mechanism has recently started to been observed in low-mass galaxies.
Investigating the presence of AGN in nearby dwarf galaxies
using mid-infrared emission, \cite{Marleau17} identified $303$ candidates,
of which $91 \%$ were subsequently confirmed as AGN by other methods.
The stellar masses of these galaxies are estimated to be between $10^{6} - 10^{9} M_{\odot}$;
and the black hole masses in the range $10^{3} - 10^{6} M_{\odot}$.
\cite{Penny17} presented observational evidence for AGN feedback in a sample of
$69$ quenched low-mass galaxies $(M_{\star} < 4 \times 10^{9} M_{\odot})$;
including $6$ galaxies showing signatures of an active AGN preventing ongoing star-formation.

The concordance $\Lambda$CDM cosmological scenario of galaxy formation presents
multiple challenges in the dwarf galaxy mass range: e.g. core-cusp, number of DGs.
Recently \cite{Silk17} made an exciting claim that the presence of IMBHs
at the centers of essentially all old Dwarf Galaxies (DGs) can potentially solve the problems.
Early feedback from these IMBHs output energy and affect the host gas-rich DGs at $z = 5 - 8$.
This early feedback can quench star-formation, reduce the number of DGs,
and impact the density profile at DG centers.
\cite{Dashyan17} studied the same problem analytically,
and compared AGN versus SN feedback. They find a critical halo mass
below which the central AGN can drive gas out of the host halo.
This negative feedback effect of AGN is found to be more efficient than SN
in the most massive DGs, where SN is not able to expel the gas. 

Here, we investigate the scenario that IMBHs are present at the centers
of all dwarf galaxies, by performing cosmological hydrodynamical simulations.
Our goals are to (i) test if IMBHs would grow at DG centers, 
and (ii) quantify the impact on star formation.

%

\begin{table}
\begin{minipage}{1.0 \linewidth}
\caption{
Simulation runs and parameters. 
}
\label{Table-Sims}
\begin{tabular}{@{}cccccc}

\hline

Run  & BH      & Min. Halo Mass for BH Seeding, & Seed BH Mass,                & BH kinetic feedback \\
name & present & $M_{\rm HaloMin} [M_{\odot}]$  & $M_{\rm BHseed} [M_{\odot}]$ & kick velocity $v_w$ (km/s) \\

\hline 

{\it SN}         & No  & -- & -- & -- & \\                               

{\it BHs2h1e6}   & Yes & $h^{-1} \times 10^{6}$ & $10^{2}$ & $2000$ \\   

{\it BHs2h7e7}   & Yes & $5 h^{-1} \times 10^{7}$ & $10^{2}$ & $2000$ \\ 

{\it BHs3h1e7}   & Yes & $1 \times 10^{7}$ & $10^{3}$ & $2000$ \\        

{\it BHs3h2e7}   & Yes & $2 \times 10^{7}$ & $10^{3}$ & $2000$ \\        

{\it BHs3h3e7}   & Yes & $3 \times 10^{7}$ & $10^{3}$ & $2000$ \\        

{\it BHs3h4e7}   & Yes & $4 \times 10^{7}$ & $10^{3}$ & $2000$ \\        

{\it BHs3h4e7v5} & Yes & $4 \times 10^{7}$ & $10^{3}$ & $5000$ \\        

{\it BHs3h5e7}   & Yes & $5 \times 10^{7}$ & $10^{3}$ & $2000$ \\        

{\it BHs4h4e7}   & Yes & $4 \times 10^{7}$ & $10^{4}$ & $2000$ \\        

\hline
\end{tabular}

\end{minipage}
\end{table}


\section{Numerical Method} 

We use a modified version of the TreePM (particle mesh) -
SPH (smoothed particle hydrodynamics) code {\sc GADGET-3} (\cite{Springel05}). 
Radiative cooling and heating is incorporated from \cite{Wiersma09a}.
Eleven element species (H, He, C, Ca, O, N, Ne, Mg, S, Si, Fe) are tracked.
Star-formation is implemented following the multiphase effective sub-resolution 
model by \cite{SH03}, and chemical enrichment from \cite{Tornatore07}. 

BHs are collisionless sink particles (of mass $M_{\rm BH}$) in our simulations. 
A BH (of initial mass $M_{\rm BHseed}$) is seeded at the center of each galaxy more 
massive than a total mass $M_{\rm HaloMin}$, which does not contain a BH already. 
We test different values of minimum halo mass and seed BH mass in the range: 
$M_{\rm HaloMin} = (10^{6} - 10^{7}) M_{\odot}$, 
and $M_{\rm BHseed} = (10^{2} - 10^{3}) M_{\odot}$. 
The sub-resolution presciptions for BH accretion and 
{\it kinetic} feedback are adopted from \cite{Barai14, Barai16}.

\subsection{Simulations} 

We perform cosmological hydrodynamical simulations of small-sized
boxes to probe dwarf galaxies at high redshifts.
The initial condition at $z = 100$ is generated using the
{\sc MUSIC}\footnote{MUSIC - Multi-scale Initial Conditions for Cosmological
Simulations: https://bitbucket.org/ohahn/music} software (\cite{Hahn11}). 
The size of the cubic cosmological volume is $(2 h^{-1} ~ {\rm Mpc})^3$ comoving.
Hence the total mass of matter (dark matter + baryons) in the box is
$6.86 \times 10^{11} h^{-1} M_{\odot}$.
We use $256^3$ dark matter and $256^3$ gas particles in the initial condition.
The dark matter particle mass is $m_{\rm DM} = 3.44 \times 10^{4} h^{-1} M_{\odot}$,
and the gas particle mass is $m_{\rm gas} = 6.43 \times 10^{3} h^{-1} M_{\odot}$.
The gravitational softening length is set to $L_{\rm soft} = 0.1 h^{-1}$ kpc comoving. 
We execute a series of $10$ simulations, with characteristics listed in Table~\ref{Table-Sims}.

\section{Results and Discussion} 

\subsection{Black Hole Accretion and Growth} 

\begin{figure}
\centering
\includegraphics[width = 0.7 \linewidth]{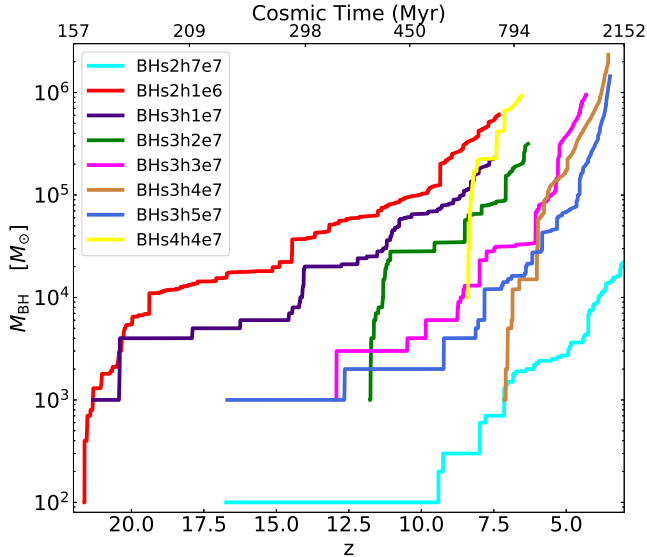}
\caption{
BH mass growth with redshift of the most-massive BH in each run. The different colours discriminate the runs.
}
\label{fig-BH-Mass-vs-Time}
\end{figure} 


We find that first BHs are seeded at different cosmic times depending on the value of
minimum halo mass for BH seeding, $M_{\rm HaloMin}$.
The seeding epoch varies between $z \sim 22$ to $z \sim 16$ in our simulations, when the first halos
reach $M_{\rm halo} = h^{-1} \times 10^{6} M_{\odot}$ to $M_{\rm halo} = 5 \times 10^{7} M_{\odot}$. 
The redshift evolution of the most-massive BH mass in the BH runs is plotted in Fig.~\ref{fig-BH-Mass-vs-Time}.
Each BH starts from an initial seed of $M_{\rm BH} = 10^{2} M_{\odot}$ in the runs named {\it BHs2*},
$10^{3} M_{\odot}$ in the runs named {\it BHs3*}, and $10^{4} M_{\odot}$ in the runs named {\it BHs4*}.
The subsequent mass growth is due to merger with other BHs (revealed as vertical rises in $M_{\rm BH}$),
and gas accretion (visualized as the positive-sloped regions of the $M_{\rm BH}$ versus $z$ curve). 
The final properties reached depends on the simulation.
The most-massive BH, considering all the runs, has grown to $M_{\rm BH} = 2 \times 10^6 M_{\odot}$
at $z = 5$ in run {\it BHs3h4e7} (brown curve in Fig.~\ref{fig-BH-Mass-vs-Time}).

\subsection{Star Formation} 

\begin{figure}
\centering
\includegraphics[width = 0.7 \linewidth]{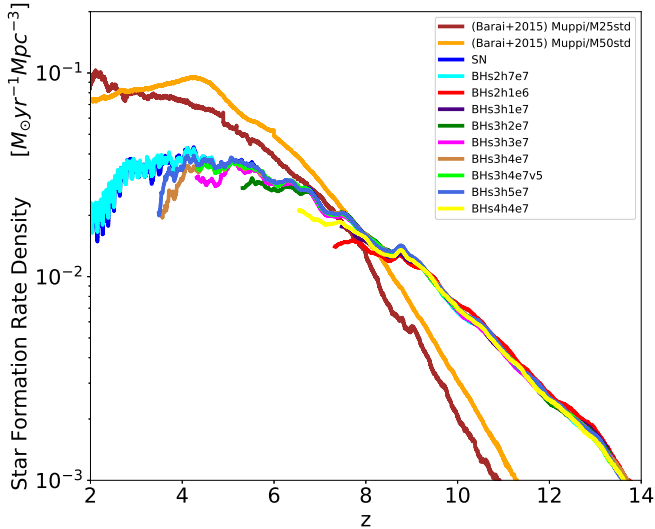}
\caption{
Total star formation rate density in simulation volume as a function of redshift.
}
\label{fig-SFR-total}
\end{figure}

The Star Formation Rate Density
(SFRD in units of $M_{\odot} yr^{-1} Mpc^{-3}$, counting stars forming in the whole simulation box)
versus redshift of the simulation runs is displayed in Fig.~\ref{fig-SFR-total}. 
The SFRD rises with time in the {\it SN} run (blue curve in Fig.~\ref{fig-SFR-total}) initially from $z \sim 15$,
reaches a peak at $z \sim 4$ (the peak epoch of star-formation activity in the Universe),
and decreases subsequently over $z \sim 4 - 2$. 
The presence of a BH quenches star formation by accreting some gas in,
ejecting some gas out of the halo as outflows, and/or heating the gas. 
The models suppress SF substantially from $z \sim 8$ onwards, when the BHs have grown massive. 
We find that BHs need to grow to $M_{\rm BH} > 10^5 M_{\odot}$,
in order to suppress star-formation, even in these dwarf galaxies.
BH feedback causes a reduction of SFR up to $5$ times in the different runs.

The red curve (run {\it BHs2h1e6}) already quenches SF as early as $z \sim 8$.
This is because the BH has already grown to $M_{\rm BH} \sim 5 \times 10^{5} M_{\odot}$ at that epoch,
more massive than all the other runs.
As another example, the brown (run {\it BHs3h4e7}) and royal-blue (run {\it BHs3h5e7}) curves
quench SF from $z \sim 4.5$ to $z \sim 3.5$.
This is the epoch when the BH masses in these runs increase from
$M_{\rm BH} = 10^{5} M_{\odot}$ to $M_{\rm BH} = 10^{6} M_{\odot}$.

\section{Conclusions} 

We conclude that
(i) IMBHs at DG centers grow from $10^{2} - 10^{3} M_{\odot}$ to $10^{5} - 10^{6} M_{\odot}$
by $z \sim 4$ in a cosmological environment.
These IMBHs in DGs can become the seeds of supermassive BHs
(which grows to $M_{\rm BH} \sim 10^9 M_{\odot}$) in massive galaxies.
(ii) Star formation is quenched when the BHs have grown to $M_{\rm BH} > 10^5 M_{\odot}$.
We find a positive correlation between the mass growth BHs and the quenching of SF.

\end{document}